\documentclass[12pt]{iopart}
\usepackage{epsf}
\usepackage{psfig}
\begin{document}
\title[Charge-ordering quantum criticality]{Charge-ordering 
quantum criticality in the phase diagram of the  cuprates}  
\author{C. Di Castro, M. Grilli, and S. Caprara}  
\address{Istituto Nazionale per la Fisica della Materia and  
Dipartimento di Fisica, Universit\`a di Roma ``La Sapienza'',\\  
Piazzale Aldo Moro 2, I-00185 Roma, Italy}  

\ead{marco.grilli@roma1.infn.it}

\begin{abstract} 
We discuss how the quantum-critical-point scenario for high-$T_c$ 
superconductors with density-driven charge order or stripe formation 
finds support from recent EXAFS experiments. 
Our phase diagram has the interesting feature, which is also suggested by
the EXAFS 
experiments, that some cuprates may not show well-formed stripe phase.
We also consider the 
extensions of the scenario to include the interesting possibility of 
first-order transitions which call for further experimental and theoretical 
investigation.  
\end{abstract}  
  
\maketitle

There are by now several evidences of the existence of a Quantum Critical  
Point (QCP)  arountd the optimal doping of the 
cuprates\cite{BOEBINGER,TALLON,PANAGOPOULOS,FENG,ZEIT}.   
The issue we address here is to interpret the recent experimental results 
of Ref. \cite{BIANCONI} as further evidence for the QCP 
scenario for high $T_c$ cuprates, that we have been proposing along the 
years\cite{RASS}. This scenario found one of its possible realizations within 
a model of strongly correlated electrons interacting with phonons, i.e. the 
simple one-band Hubbard-Holstein model in the presence of long-range 
Coulombic forces\cite{CAST,BECCA}. Starting from the comparison 
with the recent experiments, we here also 
propose possible extensions of this model. The simplifications with respect 
to models taking into account the complex structure of the cuprates was 
adopted for two main reasons: first of all it makes the model manageable 
despite the formal complications related to the treatment of strong e-e 
correlations via the slave-boson large-N expansion formalism. Moreover it 
allows to concentrate on the relevant spects providing both a non-Fermi-liquid
behavior and a strong pairing mechanism. Although the relevant features 
related to quantum criticality should be generic, different cuprates should 
of course show different aspects related to their specific structure. 
  
Two major results were obtained within the Hubbard-Holstein 
model\cite{CAST,BECCA}: 
i) Charge collective fluctuations mediate a 
doping- and temperature-dependent singular scattering among the  
quasiparticles, and ii) the $T=0$ phase diagram here reported in Fig. 1 with 
a line of the electron-phonon (e-ph) coupling $g$ as a function of the doping 
$\delta$. This line marks a second-order transition for the onset of charge 
ordering (CO) characterized by an order parameter $\rho_{q_c}$ representing 
the microscopic charge modulation at a wavevector $q_c$
\footnote{In the framework of Ref. \cite{BECCA} 
	the charge ordering
        instability is a second-order transition leading to a
        gradual increase of the charge modulation. It is only when
        one enters deeply inside the (locally) ordered phase that
        anharmonic distortions of the charge profile may arise
        from the enhanced interactions with the spin and the lattice
        degrees of freedom. This may lead to the stripe formation}. The 
low-doping region is only indicative since the magnetic effects have not been 
considered. The separation between the homogeneous phase and the CO phase is 
marked by a line having a minimum $g=g_{min}$ controlling the transition. For 
$g<g_{min}$ no CO (or stripes) could be formed as a static phase. Of course 
strong 
quantum fluctuations mediating pairing and 
non-Fermi-liquid behavior in the normal phase 
would be present both above and below the quantum critical line. For each 
value of $g_c(\delta_c)$ there is a QCP, which is the $T=0$ end-point of a 
critical line $T_{CO}(\delta)$. This line extends to finite temperature the 
effective scattering mediated by the critical fluctuations. A recent 
analysis\cite{ANDERGASSEN} has shown that $T_{CO}$ is of the order of the 
observed pseudogap crossover temperature $T^*$. Indeed, within the CO-QCP 
scenario, the pseudogap temperature $T^*$ arises because the quasiparticles 
feel an increasingly strong attractive interaction by approaching the critical 
line $T_{CO}(\delta)$. In the particle-hole channel, this interaction can 
produce a gap due to the incipient CO\cite{gapPH}. At the same time in the 
particle-particle channel, the strong attraction can lead to the formation of 
Cooper pairs even in the absence of phase coherence, which is only 
established at a 
lower temperature $T_c$\cite{gapPP}. Therefore $T^*$ closely tracks the 
underlying transition line $T_{CO}(\delta)$, and merges with $T_c$ 
around optimal doping, when the opening of the pseudogap and the phase 
coherence occur simultaneously.
\begin{figure}  
\label{FIG1}  
{\hspace {3.truecm}
\hbox{\psfig{figure=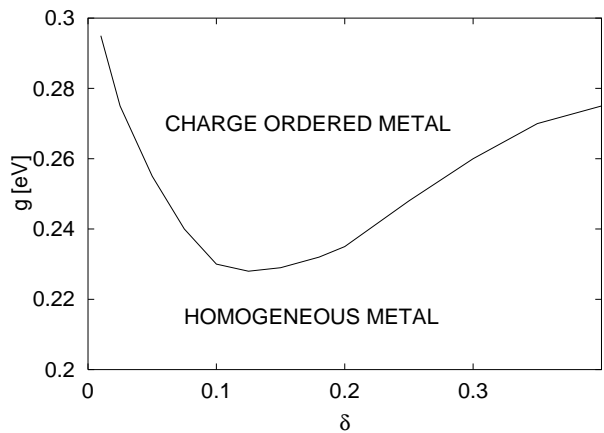,width=8.5cm,angle=-90}}}  
\caption{Phase diagram  e-ph coupling $g$ vs doping $\delta$
of the single-band infinite-$U$ Hubbard-Holstein model
with nearest-neighbor hopping $t_{\rm phys}=0.5$eV, 
next-to-nearest-neighbor hopping $t_{\rm phys}'=-(1/6)t_{\rm phys}$, phonon 
frequency
$\omega_{0\, \rm phys}=0.04$eV, and in the presence of long-range 
Coulombic forces with strength $V_{C\, \rm phys}=0.55$eV.
  See Ref. \cite{BECCA} for a detailed
description of the model.}
\end{figure}  

The Hubbard-Holstein model is therefore able to describe a system  
with a (lattice-driven) electronic instability at $T=0$ occurring  
below optimal doping, which divides the $T$ vs. $\delta$ phase  
diagram into a (nearly) ordered, quantum critical and a quantum  
disordered region naturally corresponding to the under-, optimal,  
and over-doped regions respectively. However, despite this  
gross correspondence, the model and its solution are too schematic  
to determine the relationship between the model parameters and  
the microscopic quantities of the real materials (in this sense  
the Hubbard-Holstein model can be regarded as the analog of the Ising  
model in critical phenomena: More  involved models lead
to different universality classes maintaining the main features
related to scaling). In particular,  
the specific phonons involved, their coupling to the quasiparticles,  
the orbitals needed for a proper band description, are  
open issues presently under investigation. In this regard, experiments  
can provide crucial informations. In particular, recent EXAFS 
experiments \cite{BIANCONI} identified interesting features associated to  
the lattice structure. A microstrain of the Cu-O  
bonds was measured in terms of the deviation of the Cu-O distance   
with respect to a reference distance $d_0=1.985 (\pm 0.05) \, \AA$  
\begin{equation}  
\varepsilon =2 \left(1-{R_{Cu-O}\over d_0}\right),  
\end{equation}  
which introduces a mismatch in the lattice between the CuO$_2$  
layers and the rock-salt layers.  
In the $(\varepsilon,\delta)$ plane the experiments suggest a line  
separating a homogeneous phase from a ``coexistence region  
of fluctuation bubbles of superconducting stripes''\cite{BIANCONI}. 
Actually, the   published experimental data summarized 
in Fig. 7 of Ref. \cite{BIANCONI}
leave various possibilities open:  
i) the above line could mark second-order transitions smoothly  
connecting the homogeneous and the ``charge-ordered'' phases,  
alternatively ii) it could be a first-order line ending into a  
single QCP or into a finite segment of QCP's, or, finally,  
iii) it could entirely be a line of first-order transitions.  
The experiments also suggest the remarkable possibility,   
 that some cuprates may not show a stripe phase since their  
microstrain is lower than a critical value $\varepsilon_c$  
estimated in Ref. \cite{BIANCONI}
to be $\varepsilon_c\approx 0.045$ from the onset of local 
lattice distortions.
  
We address now the issue whether and how the  
experimental results presented in Ref. \cite{BIANCONI}
can be connected with  the above theoretical model and its extensions,
both at the level of the  
calculated properties and of the open possibilities   
to be investigated with a further analysis.
First of all one observes that the close resemblance between the  
experimental $(\varepsilon, \delta)$  and the  
theoretical $(g,\delta)$ phase  diagrams is a clear indication that  
the Cu-O microstrain and the e-ph coupling $g$ of the simplified   
Hubbard-Holstein model are strictly related. From the microscopic  
point of view it is indeed quite natural that a lattice contraction  
in the CuO$_2$ planes can enhance the effective coupling between  
the electrons and the ions. Therefore the comparison between experiments and  
theory allows to draw the conclusion that a one-to-one monotonic  
relation $g=g(\varepsilon)$ is likely to exist.   
  
Once this general framework is settled, one can move to   
identify more detailed possible scenarios. In particular  
the order of the homogeneous-metal to charge-ordered-metal  
transition is of obvious relevance. The simplest possibility  
is that in the real materials this transition is of the second order.  
In this case, at $T=0$,  the different microstrains  
determined by the rock-salt layers directly correspond  
to different e-ph couplings and are reflected in the  
similar phase diagrams. Of course this simple picture  
is not necessarily realized in the quite complex real materials  
where i) anharmonic effects and/or ii) additional non-ordering  
fields can partially or entirely transform the second-order transition  
into a first-order one. Both  
these possibilities can be considered within our Hubbard-Holstein  
scheme. The idea, which is standard in the theory of   
critical phenomena\cite{IMRY},   
 is that both these mechanisms can change  
the sign of the quartic term $u|\rho_{q_c}|^4$ of an effective Ginzburg-Landau 
description of the free energy of the system.   
If this is the case a first-order transition is obtained and
higher-order terms like $v|\rho_{q_c}|^6$ become  
important\cite{IMRY}.  
In case i) diagrams like in Fig. 2a, involving a coupling between  
a first and a second harmonic charge modulation  
${\tilde u}(\rho_{q_c}^2\rho_{-2q_c}+c.c.)$,  
induce a negative correction to $u$. 
\begin{figure}
\label{FIG2}    
{ \hspace {3.5 truecm}
\hbox{\psfig{figure=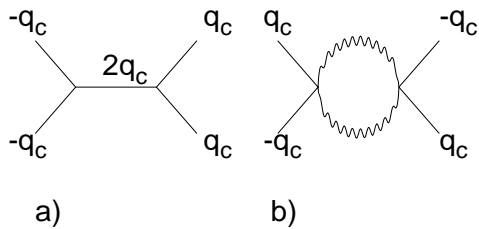,width=8.5cm,angle=-90}}
}  
\caption{Second-order diagrams correcting the quartic
term in the ordering field $\rho_{q_c}$. a) exemple of
contribution  from anharmonic fluctuations, and b) exemple
of contribution from fluctuations of a non-ordering field
(represented by the wavy lines).}  
\end{figure}  

If the charge fluctuations around the second-harmonic $2q_c$ are large   
enough, this correction $(\propto -{\tilde u}^2)$  
can drive $u$ to negative values.  
The alternative possibility ii) is that the microstrain   
enhances the coupling of the electrons with an additional  
phonon field, in general not coupled {\it \`a la }
Holstein, but, more likely, coupled to the Cu-O hopping
like in Su-Schrieffer-Heeger models. At small values of this   
coupling the fluctuations of the additional phonon field   
weakly dress the electrons, which still undergo a second-order  
instability because of the coupling to the original phonon mode.  
However, for sufficiently large values of the microstrain, it could  
happen that the additional phonon field introduces a 
non-ordering field $\psi$, which is generically bilinear in
the electronic variables. Within the Ginzburg-Landau approach,
this generates a four-leg vertex of the form
$u'|\rho_{q_c}|^2|\psi|^2$.  
Second-order processes $(\sim -{u'}^2)$ involving the
propagator of the new field $\psi$,
like the one depicted in Fig. 2b, can also drive the $u|\rho_{q_c}|^4$ term  
negative and naturally transform partially or entirely the   
line of QCP's in Fig. 1 into a line of first-order transition.
According to the grneral discussion of density-driven first-order phase
transitions in the presence of long-range forces\cite{JOSE}, nearby the
first-order line one expects in this case bubbles of stripes in the 
homogeneous metallic background.

We stress here that these two possible mechanisms to change the  
order of the CO transition naturally extend  
the simple second-order transition theoretically determined   
in Ref. \cite{BECCA}. In the previous work the technical   
large-N scheme within the leading-order approximation   
and the consideration of a single phonon mode  
did not allow for the inclusion of processes like  
those exemplified in Fig. 2. Nevertheless, the above-mentioned  
resemblance of the experimental and the theoretical phase  
diagrams make it clear that the qualitative determination of  
the region where the instability takes place is already  
captured within the simplified scheme. Without qualitatively modifying
the generic appearance of the phase diagram, 
the additional terms discussed above would simply shift the 
instability line and possibly replace partially or completely
the second-order line with first-order coexistence regions.

We also notice that the reentrant behavior of the instability
region at low doping is a consequence of the increased
residual repulsion between the quasiparticles at finite
wavelengths, when the large quasiparticle effective mass
no longer allows for an effective screening of the large (actually
infinite in our model) Hubbard repulsion between the electrons.
This physically reasonable effect occurs regardless of 
magnetism, which is surely present in real systems at low doping, but
is poorly described within out large-N approach. Nevertheless,
if the resemblance between our theoretical and the
experimental phase diagrams is confirmed, this  suggests that
magnetic effects should become relevant at doping lower than
the doping at which the reentrant behavior sets in.

Further experiments are needed to establish which of the above-mentioned 
possibilities i)-iii) for the phase diagram is realized in the 
cuprates. However, in all the three
 cases the implications of our QCP scenario maintain  
their full validity unless the transition becomes
strongly first-order. Indeed all the relevant observables 
mainly depend on the deviation from criticality fixed by 
$\xi^{-2}(T, \delta-\delta_c)$, where $\xi$ is the  
correlation length for the onset  
of charge ordering. The only difference is now that  
$\xi$ would acquire an additional dependence on $\varepsilon-\varepsilon_c$.  
We stressed in several papers\cite{RASS} 
that the actual onset of a fully   
developed CO phase was competing with local or coherent  
pair formation,  which modifies the fermionic spectrum  
stabilizing the system against the electronic CO
transition. There is the complementary possibility  
that the CO state can never be reached because $g$ is below $g_{min}$. The 
corresponding observable quantity identified by the experiments in  
Ref. \cite{BIANCONI} is the critical microstrain $\varepsilon_c$,
below which no stripe phase can be observed.
 Therefore the cuprates can deviate  
from criticality not only by $\delta-\delta_c$ in the (overdoped)  
quantum-disordered region, or by $T$ and $T-T_{CO}(\delta)$ in the  
quantum-critical or underdoped region respectively, but also because 
$\varepsilon$ is smaller or larger than $\varepsilon_c$, thereby tending to  
a homogeneous or an inhomogeneous phase respectively. As long as   
$\varepsilon-\varepsilon_c$ is not too large and
$T$ is finite, for $\delta\sim \delta_c$
we enter again the quantum-critical   
region with strong dynamical fluctuations in both cases. 
According to the proposed mechanism of pairing
mediated by critical fluctuations, the stronger are these
fluctuations and the larger is $T_c$. Indeed in the experiment\cite{BIANCONI}
the mercury compound Hg1212, having  $\varepsilon \sim \varepsilon_c$, 
has the the largest $T_c$, while Hg1201 (with
$\varepsilon < \varepsilon_c$) and Bi2212 or La214 
(with $\varepsilon > \varepsilon_c$) have a lower $T_c$.
Moreover, since $\varepsilon < \varepsilon_c$ in Hg1201, we expect
no $T_{CO}(\delta)$, and therefore no $T^*$, for this material.
  
\ack
We acknowledge stimulating discussions with C. Castellani.

\end{document}